\documentclass[11pt]{article}

\usepackage{amsmath}
\usepackage{graphicx}
\usepackage{amsfonts}
\usepackage{amssymb}
\usepackage{epsfig}
\usepackage{color}
\usepackage{psfrag}
\usepackage{epstopdf}
\usepackage{caption}
\usepackage{subcaption}

\newcommand{\EQ}{\begin{equation}}
\newcommand{\EN}{\end{equation}}
\newcommand{\be}{\begin{equation}}
\newcommand{\ee}{\end{equation}}
\newcommand{\bea}{\begin{eqnarray}}
\newcommand{\eea}{\end{eqnarray}}

\newcommand{\goto}{\rightarrow}

\def\goto{\longrightarrow}

\newcommand{\nn}{\nonumber \\}

\begin{document} \setcounter{page}{0}
\topmargin 0pt
\oddsidemargin 5mm
\renewcommand{\thefootnote}{\arabic{footnote}}
\newpage
\setcounter{page}{0}
\topmargin 0pt
\oddsidemargin 5mm
\renewcommand{\thefootnote}{\arabic{footnote}}
\newpage

\begin{titlepage}
\begin{flushright}
%SISSA 40/2012/EP \\
%DFTT 9/2007
\end{flushright}
\vspace{0.5cm}
\begin{center}
{\large {\bf On the $RP^{N-1}$ and $CP^{N-1}$ universality classes}}\\
\vspace{1.8cm}
{\large Youness Diouane$^{1,2}$, Noel Lamsen$^{3}$ and Gesualdo Delfino$^{1}$}\\
\vspace{0.5cm}
{\em $^{1}$SISSA and INFN -- Via Bonomea 265, 34136 Trieste, Italy}\\
{\em $^{2}$ICTP, Strada Costiera 11, 34151 Trieste, Italy}\\
{\em $^3$National Institute of Physics, University of the Philippines Diliman,\\ 1101 Quezon City, Philippines}\\
%{\em $^2$INFN sezione di Trieste}\\
%{\em E-mail: delfino@sissa.it}\\
%\vspace{0.5cm}
%{\large and}\\
%\vspace{0.5cm}
%{\large xyz}\\
%\vspace{0.5cm}
%{\em Department of Physics, University of}}

\end{center}
\vspace{1.2cm}

\renewcommand{\thefootnote}{\arabic{footnote}}
\setcounter{footnote}{0}

\begin{abstract}
\noindent
We recently determined the exact fixed point equations and the spaces of solutions of the two-dimensional $RP^{N-1}$ and $CP^{N-1}$ models using scale invariant scattering theory. Here we discuss subtleties hidden in some solutions and related to the difference between ferromagnetic and antiferromagnetic interaction.
\end{abstract}
\end{titlepage}

\newpage
%\tableofcontents

\section{Introduction}
Symmetry plays a fundamental role in the theory of critical phenomena. The common case is that of criticality induced by the spontaneous breaking of a global symmetry, which also controls the universality class the statistical system falls into. Less understood has been the effect that additional local symmetries of the Hamiltonian may induce. The basic case study has been provided by the lattice $RP^{N-1}$ model which, besides a global $O(N)$ symmetry, displays invariance under local spin reversal, thus realizing the head-tail symmetry characteristic of liquid crystals \cite{deGP}. While in three dimensions the $RP^{N-1}$ ferromagnet has a first order transition \cite{ZMZ}, the two-dimensional case has offered several reasons of interest. In principle this case could provide important insight through the exact methods of lattice integrability \cite{Baxter,Nienhuis_review} and conformal field theory \cite{BPZ,DfMS}, but the model traditionally remained inaccessible to them. It has then been the object of many numerical studies, which however found difficult to reach conclusions, especially due to the very large correlation length in the low temperature region of main interest. It was proposed that, in absence of spontaneous breaking of continuous symmetries \cite{MWHC}, finite temperature criticality may be produced by a topological transition of the Berezinskii-Kosterlitz-Thouless (BKT) type \cite{BKT} mediated by disclination defects \cite{Stein,Mermin}. While the transition should definitely occur for $RP^1\sim O(2)$, its existence for $N>2$ has been debated in numerical studies \cite{CPZ,KZ,FPB,DR,PFBo,FBBP,Tomita,SGR,KS}. The most likely alternative to a topological transition is that criticality is limited to zero temperature, and numerical investigations tried to establish if the local symmetry affects the universality class \cite{Sinclair,CEPS,CEPS2,NWS,Hasenbusch,CHHR,BFPV}. 

In \cite{DDL} we determined for the first time the exact fixed point equations of the $RP^{N-1}$ model in two dimensions. This was achieved in the framework of scale invariant scattering theory \cite{paraf} (see \cite{sis} for a review) which in recent years provided new information \cite{random,DT1,DT2,DL_ON_jhep,DL_ON,DL_vector_scalar,DL_softening,potts_qr} on difficult problems of two-dimensional criticality, including quenched disorder. The space of solutions of the $RP^{N-1}$ fixed point equations, containing both ferromagnetic and antiferromagnetic fixed points, was investigated in \cite{DDL,RPN,CPN} for continuous positive values of $N$ and revealed the following properties:

{\bf a.} the space of solutions corresponds to an order parameter with $M_N=\frac{1}{2}N(N+1)-1$ components;

{\bf b.} quasi-long-range order is limited to $N=2$, and there is no evidence of a topological transition above this value;

{\bf c.} for $N\leq N_*=2.24421..$ the space of solutions contains several branches of fixed points which can be relevant for gases of intersecting loops;

{\bf d.} for $N>N_*$ there is only one scattering solution;

{\bf e.} the scattering solution of the previous point describes a zero temperature fixed point in the $O(M_N)$ universality class, with central charge $M_N-1$.

After our papers, the $RP^2$ model was numerically investigated in \cite{UO,BDVVV} using the tensor network renormalization (TNR) method, which has access to the central charge and the degeneracies in the spectrum of scaling dimensions. The conclusions of these studies are consistent with our exact results up, apparently, to one finding of \cite{UO}. More precisely, while seeing signatures of a crossover at finite temperature, both studies concluded for the absence of a true topological transition, in agreement with our point {\bf b}. The authors of \cite{UO} also investigated the zero temperature behavior and saw criticality with the most relevant operator possessing $5$ components, in agreement with our point {\bf a}. For the central charge of the $RP^2$ antiferromagnet at $T=0$ they found a value consistent with $4$, in agreement with our point {\bf e}. For the ferromagnet at $T=0$, however, they found central charge $2$. They suggested that this latter fixed point could correspond to an extra scattering solution pointed out in \cite{DDL,RPN} and existing only for $N=3$. Meanwhile, however, we had shown in \cite{CPN} that this solution is only an alternative realization of the scattering matrix of point {\bf d}. It would then appear that the exact fixed point equations are missing the ferromagnetic fixed point. Here we will argue that this is not the case and that the above points {\bf a}-{\bf e} should be complemented with the following point

{\bf f.} the scattering solution of point {\bf d} also describes a zero temperature fixed point in the $RP^{N-1}$ universality class, with central charge $N-1$. 

\noindent
The fact that a single critical scattering solution can correspond to different universality classes is not new and had been illustrated in \cite{DT1} for the Potts model. In the present case it is related to the features of asymptotic freedom.

In \cite{CPN} we determined the exact fixed point equations of the $CP^{N-1}$ model, which provides the basic lattice realization of a continuous local symmetry. For their space of solutions we found properties that can again be stated as in points {\bf a}-{\bf e} above, provided we define new values $M_N=N^2-1$ and $N_*=2$, and in point {\bf b} we substitute $N=2$ with $N=\sqrt{3}$. Here we will argue that also in this case these findings should be complemented by point {\bf f}, which now will refer to the $CP^{N-1}$ universality class with central charge $2(N-1)$.  

The paper is organized as follows. In the next section we recall some generalities about symmetry, central charge and scattering. The $RP^{N-1}$ and $CP^{N-1}$ models are then discussed in sections~\ref{real} and \ref{complex}, respectively, while section~\ref{final} contains some final remarks.

\section{Symmetry, central charge and scattering}
A universality class of critical behavior is identified by a renormalization group fixed point with its specific field content. This field content, through the associated operator product expansion (OPE), implements the internal symmetry of the fixed point Hamiltonian. The field with lowest scaling dimension -- i.e. the most relevant field, in the renormalization group sense -- carrying a representation of the internal symmetry is the order parameter field $s(x)$, while the most relevant field (excluding the identity) invariant under the symmetry transformations is the energy density field $\varepsilon(x)$. Adding to the fixed point Hamiltonian the contribution of $\varepsilon$ moves the system away from the critical temperature $T_c$. The scaling dimensions $X_s$ and $X_\varepsilon$ of the order parameter and energy density fields determine the canonical critical exponents (see e.g. \cite{Cardy_book}). 

Symmetry alone does not identify a fixed point. For example, a system with a given symmetry can also exhibit a tricritical point, which possesses an additional relevant symmetry-invariant field $\varepsilon'$, and a field content larger than that of the critical point. In two dimensions a main parameter which grows with the size of the space of fields is the central charge $c$ of a fixed point. A basic illustration is obtained recalling that at critical points of statistical systems scale invariance is promoted to conformal invariance \cite{DfMS}, whose simplest realizations are the minimal models \cite{BPZ} with central charge 
\EQ
c=1-\frac{6}{p(p+1)}\,,\hspace{1cm}p=3,4,\ldots\,,
\label{c_p}
\EN
and scalar primary fields with scaling dimensions
\EQ
X_{m,n}=\frac{[(p+1)m-pn]^2-1}{2p(p+1)}\,,\hspace{.7cm}m=1,2,\ldots,p-1\,,\hspace{.4cm}n=1,2,\ldots,p\,.
\label{delta_mn}
\EN
The central charge grows with $p$, and then with the number of primary fields\footnote{The role of the central charge in establishing a hierarchy of fixed points is illustrated by the $c$-theorem \cite{cth}.}, each of which possesses infinitely many "descendants" with scaling dimensions exceeding by integers that of the primary. The critical point of the Ising model corresponds to $p=3$, while more generally the above minimal models describe multicriticality of order $(p-1)$ in $\mathbb{Z}_2$-symmetric systems. 

In general, on the other hand, the central charge does not identify a universality class, since it does not uniquely specify the field content. The simplest illustration is provided by the minimal model with $p=5$, which describes both the tetracritical $\mathbb{Z}_2$-symmetric point and the critical three-state Potts model, which has permutational symmetry $\mathbb{S}_3$. While the $\mathbb{Z}_2$-symmetric tetracritical point possesses all the primary dimensions (\ref{delta_mn}) with multiplicity one \cite{BPZ}, the Potts critical point only possesses a subset of them, with a field-dependent multiplicity \cite{Dotsenko,Cardy_modular}. In particular, $X_s=X_{2,2}$ and $X_\varepsilon=X_{3,3}$ for $\mathbb{Z}_2$, while $X_s=X_{2,3}$ and $X_\varepsilon=X_{2,1}$ for Potts; in the latter case $X_{2,3}$ has multiplicity two, as required for the $\mathbb{S}_3$ order parameter.  

When looking for fixed points of the renormalization group in the scattering framework \cite{paraf,sis}, one of the Euclidean dimensions of the statistical system is taken as imaginary time and conformal invariance is implemented in the basis of the massless particle excitations. It is crucial that in two dimensions conformal symmetry has infinitely many generators \cite{DfMS}, which provide infinitely many quantities that have to be conserved in a scattering process. As a result, the initial and final states are kinematically identical (complete elasticity) and the scattering problem greatly simplifies. An additional simplification is that scale invariance leads to two-particle amplitudes which do not depend on the center of mass energy, the only relativistic invariant in ($1+1$)-dimensional scattering. All this results in a particularly simple form of the crossing and unitarity equations which constrain any relativistic scattering theory \cite{ELOP}. We call these equations the fixed point equations since, by construction, their solutions correspond to fixed points of the renormalization group. Since no approximation is involved in the derivation, the equations are exact. 

Within the scattering framework, the information about the universality class comes from internal symmetry, which determines the particle content and the allowed processes. Since we recalled how symmetry alone does not identify a fixed point, it follows that in general the fixed point equations do not possess a single solution, but rather a space of solutions containing the different fixed points with the given symmetry. This space of solutions will contain critical and multicritical points, as well as the fixed points corresponding to ferromagnetic and antiferromagnetic interactions. In particular, there may be solutions for which some of the scattering amplitudes vanish, making possible fixed points with a symmetry larger than that common to the whole space of solutions. These features have been illustrated in \cite{DT1} for the $q$-state Potts model and in \cite{DL_ON} for the $O(N)$ model, and are reviewed in \cite{sis}.

Another possibility, more subtle to identify, is that a single scattering solution describes different universality classes. An explicit example is provided by the $q$-state Potts model \cite{Wu} with $q\in[0,4]$ which can be considered as a continuous variable and parametrized as $\sqrt{q}=2\cos(\pi/p)$. The ferromagnetic model possesses, besides the fundamental critical line, a tricritical line with central charge (\ref{c_p}) \cite{DF}. On the other hand, the antiferromagnetic model on the square lattice possesses a critical line\footnote{This antiferromagnetic line corresponds to nonnegative temperature only up to $q=3$ \cite{Baxter_AF}.} with central charge $c=2(p-3)/p$ \cite{Saleur}. When the scale invariant scattering solutions are obtained implementing the permutational symmetry $\mathbb{S}_q$ characteristic of the Potts model, these two critical lines correspond to the same solution \cite{DT1}. The mechanism allowing this is conveniently illustrated considering the amplitude of the symmetry invariant scattering channel, which quite generally can be written as \cite{paraf,sis}
\EQ
S=e^{-2i\pi\Delta_\eta}\,,
\label{S_phase}
\EN
where $\Delta_\eta$ is the conformal dimension of the chiral field which creates the particles at criticality. Its value is $\Delta_\eta=X_{3,1}/2=1+2/p$ along the ferromagnetic tricritical line, and $\Delta_\eta=2/p$ along the square lattice antiferromagnetic critical line \cite{DT1}. Since the two values differ by an integer, they give the same amplitude (\ref{S_phase}).

In the following we will see the relevance of these considerations for the $RP^{N-1}$ and $CP^{N-1}$ models, whose spaces of critical scattering solutions were obtained in \cite{DDL,RPN,CPN}.

\section{$RP^{N-1}$ model}
\label{real}
\subsection{Scale invariant scattering}
The $RP^{N-1}$ lattice model is  defined by the Hamiltonian
\EQ
{\cal H}_{RP^{N-1}}=-J\sum_{\langle i,j\rangle}({\bf s}_i\cdot{\bf s}_j)^2\,,
\label{lattice}
\EN
where ${\bf s}_i=(s_{1,i},s_{2,i},\ldots,s_{N,i})$ is a $N$-component unit vector located at site $i$. The interaction is ferromagnetic for $J>0$ and antiferromagnetic for $J<0$. The Hamiltonian differs from that of the $O(N)$ vector model for the square in the r.h.s., which makes (\ref{lattice}) invariant also under local inversions ${\bf s}_i\to -{\bf s}_i$ (head-tail symmetry). It follows that ${\bf s}_i$ effectively takes values on the unit hypersphere with antipodal points identified, namely the $RP^{N-1}$ manifold. This is taken into account by the order parameter variable \cite{deGP}
\EQ
Q_{ab,i}=s_{a,i} s_{b,i}-\frac{1}{N}\delta_{ab}\,.
\label{op}
\EN

\begin{figure}
\begin{center}
\includegraphics[width=15cm]{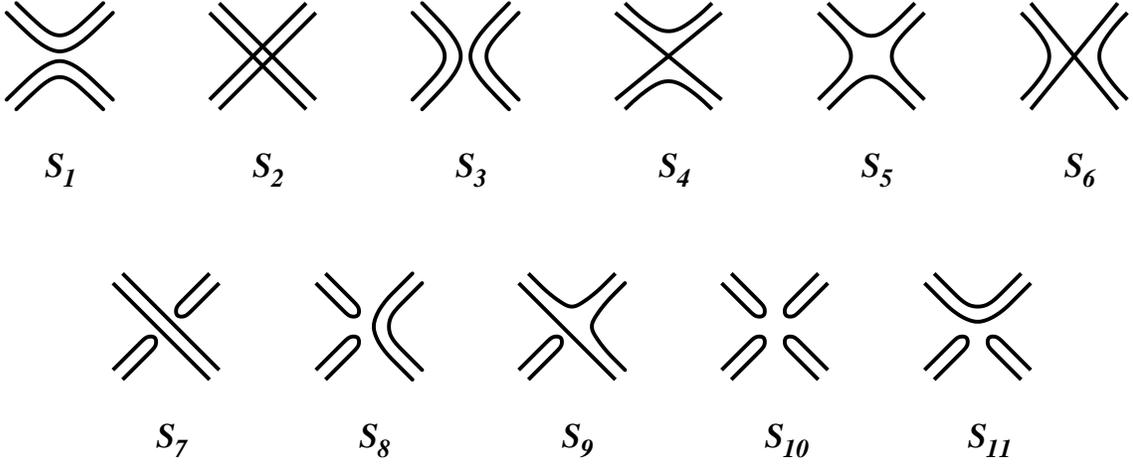}
\caption{Scattering amplitudes entering (\ref{S_tensor}). Time runs upwards.
}
\label{tensor_ampl}
\end{center} 
\end{figure}

The derivation of the $RP^{N-1}$ fixed points equations within the scattering framework has been performed in detail in \cite{DDL,RPN}; here we simply recall the main steps and results. In the continuum limit, the order parameter field is the traceless symmetric tensor $Q_{ab}(x)$. At criticality the massless particle excitations transform under the symmetries as the order parameter, and are labeled by a double index $ab$, with $a$ and $b$ going from 1 to $N$. The scattering processes for these particles are shown in figure~\ref{tensor_ampl}, where each terminal corresponds to an index and each line connects equal indices. The scattering matrix for particles $ab$ and $cd$ in the initial state and particles $ef$ and $gh$ in the final state is expressed in terms of the amplitudes $S_1,\ldots,S_{11}$ as
\begin{equation}
\begin{split}
S_{ab, cd}^{ef,gh} &= S_1\,\delta_{(ab),(cd)}^{(2)}\delta_{(ef),(gh)}^{(2)} + S_2\,\delta_{(ab), (ef)}^{(2)} \delta_{(cd),(gh)}^{(2)} + S_3\,\delta_{(ab),(gh)}^{(2)}\delta_{(cd),(ef)}^{(2)}\\
& + S_4\,\delta_{(ab)(gh),(cd)(ef)}^{(4)}  + S_5\,\delta_{(ab)(ef),(cd)(gh)}^{(4)} + S_6\,\delta_{(ab)(cd),(ef)(gh)}^{(4)}\\
& +S_7\left[\delta_{ab}\delta_{ef}\delta_{(cd),(gh)}^{(2)}+\delta_{cd}\delta_{gh}\delta_{(ab),(ef)}^{(2)}\right]  +S_8\left[\delta_{ab}\delta_{gh}\delta_{(cd),(ef)}^{(2)}+\delta_{cd}\delta_{ef}\delta_{(ab),(gh)}^{(2)}\right]\\
& +S_9\left[\delta_{ab}\delta_{(cd),(ef),(gh)}^{(3)}+\delta_{cd}\delta_{(ab),(ef),(gh)}^{(3)}+\delta_{ef}\delta_{(cd),(ab),(gh)}^{(3)}+\delta_{gh}\delta_{(cd),(ef),(ab)}^{(3)}\right]\\
& + S_{10}\,\delta_{ab}\delta_{cd}\delta_{ef}\delta_{gh}+S_{11}\left[\delta_{ab}\delta_{cd}\delta_{(ef),(gh)}^2+\delta_{ef}\delta_{gh}\delta_{(ab),(cd)}^{(2)}\right],
\end{split}
\label{S_tensor}
\end{equation}
where we introduced the notations
\begin{align}
\delta^{(2)}_{(ab),(cd)} & \equiv(\delta_{ac} \delta_{bd} + \delta_{ad} \delta_{bc})/2\,,
\label{delta2}\\
\label{delta3}\delta^{(3)}_{(ab),(cd),(ef)} &\equiv(\delta_{af}\delta_{bd}\delta_{ce} + \delta_{ad}\delta_{bf}\delta_{ce}+\delta_{ae}\delta_{bd}\delta_{cf} + \delta_{ad}\delta_{be}\delta_{cf} \nonumber\\ 
&+ \delta_{af}\delta_{bc}\delta_{de} + \delta _{ac} \delta_{bf} \delta_{de}+\delta _{ae} \delta _{bc} \delta _{df}+\delta _{ac} \delta _{be} \delta_{df})/8\,,\\
\label{delta4}\delta^{(4)}_{(ab)(cd),(ef)(gh)} &\equiv (\delta _{ah} \delta _{bf} \delta _{cg} \delta _{de}+\delta _{af} \delta _{bh}
   \delta _{cg} \delta _{de}+\delta _{ag} \delta _{bf} \delta _{ch} \delta_{de}+\delta _{af} \delta _{bg} \delta _{ch} \delta _{de}\nonumber \\
   &+\delta _{ah} \delta_{be} \delta _{c,g} \delta _{df}+\delta _{a,e} \delta _{bh} \delta _{cg} \delta_{df}+\delta_{ag} \delta _{be} \delta _{ch} \delta _{df}+\delta _{ae} \delta_{bg} \delta _{ch} \delta _{df} \nonumber\\ 
   &+\delta _{ah} \delta _{bf} \delta _{ce} \delta_{dg}+\delta _{af} \delta _{bh} \delta _{ce} \delta_{dg}+\delta_{ah} \delta_{be} \delta _{cf} \delta _{dg}+\delta _{ae} \delta _{bh} \delta _{cf} \delta_{dg}\nonumber \\ 
   &+\delta _{ag} \delta _{bf} \delta _{ce} \delta _{dh}+\delta _{af} \delta_{bg} \delta _{ce} \delta _{dh}+\delta _{ag} \delta _{be} \delta _{cf} \delta_{dh}+\delta _{ae} \delta _{bg} \delta _{cf} \delta _{dh})/4 \,
\end{align}
to take into account the different possible connections of the particle indices for a given process in figure~\ref{tensor_ampl}. Crossing symmetry allows the following parametrizations of the amplitudes
\bea
S_1=S_3^{*} &\equiv &  \rho_{1}\,e^{i\phi}, 
\label{cr1}\\
S_2 = S_2^* &\equiv & \rho_2,
\label{cr2}\\
S_4=S_6^* &\equiv & \rho_4 e^{i\theta}\,, \\
S_5=S_5^* &\equiv & \rho_5\,,\\
S_7=S_7^* &\equiv & \rho_7\,, \\
S_8=S_{11}^* &\equiv & \rho_8 e^{i\psi}\,,\\
S_9=S_9^* &\equiv & \rho_9\,,\\
S_{10}=S_{10}^* &\equiv & \rho_{10}\,.
\label{cr_last}
\eea
The fact that the field $Q_{ab}(x)$ that creates the particles is traceless is taken into account  defining ${\cal T}=\sum_a aa$ and requiring
\EQ
\mathbb{S}|(ab){\cal T}\rangle=S_0|(ab){\cal T}\rangle\,, \hspace{1cm}S_0=\pm 1
\label{decoupling}
\EN
for any particle state $|(ab)\rangle=|ab\rangle+|ba\rangle$. This means that the trace mode ${\cal T}$ is a decoupled particle that can be discarded; the sign factor $S_0$ takes into account that the trace mode can decouple as a free boson or a free fermion. The decoupling condition provides the relations
\begin{align}
&& S_2 + S_9 + NS_7 -S_0=
S_1 + S_9 + NS_{11}= S_3 + S_9 + NS_{8}= \nonumber\\
&& 4(S_4 + S_5 + S_6) + NS_9=
S_7 + S_8 + S_{11} + NS_{10}= 0\,,\label{decouplingconditions}
\end{align}
which are used to express the amplitudes $S_{i\geq 7}$ in terms of $S_{i\leq 6}$, namely
\begin{align}
\rho_7 &= -\frac{1}{N}(\rho_2 -S_0) + \frac{4}{N^2}(2\rho_4 \cos \theta + \rho_5) \label{sub1},\\
\rho_8\cos \psi &= -\frac{1}{N}\rho_1 \cos \phi + \frac{4}{N^2}(2\rho_4 \cos \theta + \rho_5) \label{sub2}, \\
\rho_8 \sin \psi &= \frac{1}{N}\rho_1 \sin \phi \label{sub3}, \\
\rho_9 &= -\frac{4}{N}(2\rho_4 \cos \theta + \rho_5) \label{sub4}, \\
\rho_{10} &= \frac{1}{N^2} \Big ( 2 \rho_1 \cos \phi + \rho_2 -S_0 -\frac{12}{N}(2 \rho_4 \cos \theta + \rho_5) \Big ). \label{sub5}
\end{align}
With this information, the requirement of unitarity of the scattering matrix yields the $RP^{N-1}$ fixed point equations in the form \cite{DDL}
\begin{align}
1&=\rho_1^2+\rho_2^2+4\rho_4^2\,, \label{uni1}\\
0&=2\rho_1\rho_2\cos\phi+4\rho_4^2\,,\label{uni2}\\
0&=M_N\rho_1^2+2\rho_1^2\cos2\phi+2\rho_1\rho_2\cos\phi+4\left(1-\frac{2}{N}+N\right)\rho_1\rho_4\cos(\phi-\theta)\nonumber\\
&\qquad+4\left(1-\frac{2}{N}\right)\rho_1\rho_4\cos(\phi+\theta)+\frac{32}{N^2}\rho_4^2\cos2\theta+4\left(1-\frac{2}{N}+N\right)\rho_1\rho_5\cos\phi\nonumber \\
&\qquad+ 8\left(1+\frac{8}{N^2}\right)\rho_4\rho_5\cos\theta+ 4\left(1+\frac{8}{N^2}\right)\rho_4^2+ 4\left(1+\frac{4}{N^2}\right)\rho_5^2\,,\label{uni3}\\
0&=2\rho_2\rho_5+2\rho_1\rho_4\cos(\phi+\theta)-\frac{8}{N}\rho_4^2+2\left(1-\frac{4}{N}\right)\rho_4^2\cos2\theta\nonumber\\
&\qquad+2\left(3-\frac{8}{N}+N\right)\rho_4\rho_5\cos\theta-\frac{4}{N}\rho_5^2\,,\label{uni4} \\
0&=2\rho_2\rho_4\cos\theta+\left(2-\frac{8}{N}+N\right)\rho_4^2+2\left(1-\frac{4}{N}\right)\rho_4^2\cos2\theta+2\rho_1\rho_5\cos\phi \nonumber\\
&\qquad+2\left(1-\frac{8}{N}\right)\rho_4\rho_5\cos\theta+\left(2-\frac{4}{N}+N\right)\rho_5^2\,, \label{uni5} \\
0&=2\rho_1\rho_4\cos(\phi-\theta)+2\rho_2\rho_4\cos\theta+2\rho_4^2\,, \label{uni6}
\end{align}
where
\EQ
M_N=\frac{1}{2}N(N+1)-1\,
\label{M_N}
\EN
coincides with the number of independent components of the order parameter variable (\ref{op}).

\begin{table}
\hspace{-.6cm}
{\small
\begin{tabular}{|c|c|c|c|c|c|c|c|}
\hline 
Solution & $N$ & $\rho_1$ & $\rho_2$ & $\cos\phi$ & $\rho_4$ & $\rho_5$  & $\cos\theta$ \\
\hline\hline
A1$_\pm$ & $(-\infty,\infty)$ & $0$ & $\pm 1$ & $-$ & $0$ & $0$ & $-$ \\ 
A2$_\pm$ & $[-3,2]$ & $1$ & $0$ & $\pm\frac{1}{2}\sqrt{2-M_N}$ & $0$ & $0$ & $-$ \\
A3 & $-3,2$ & $\sqrt{1-\rho_2^2}$ & $[-1,1]$ & $0$ & $0$ & $0$ & $-$\\
B1 & $2$ & $\frac{1-\rho_2^2}{\sqrt{1+3\rho_2^2}}$ & $[-1,1]$ & $-\frac{2\rho_2}{\sqrt{1+3\rho_2^2}}$ & $|\rho_2|\sqrt{\frac{1-\rho_2^2}{1+3\rho_2^2}}$ & $\frac{\rho_2(1-\rho_2^2)}{1+3\rho_2^2}$ & $-\text{sgn}(\rho_2)\sqrt{\frac{1-\rho_2^2}{1+3\rho_2^2}}$\\
B$2_\pm$ & $2$ & $\sqrt{1+2x\rho_2-\rho_2^2}$ & $\alpha_\pm(x)$ & $\frac{x}{\sqrt{1+2x\rho_2-\rho_2^2}}$ & $\sqrt{\frac{-x\rho_2}{2}}$ & $\frac{-x}{2}$ & $\frac{x+2\rho_2}{2\sqrt{-2x\rho_2}}$ \\
%B$3_\pm$ & $3$ & $\frac{2}{3}$ & $\pm\frac{1}{3}$ & $\mp1$ & $\frac{1}{3}$ & $\pm\frac{1}{3}$ & $\pm 1$\\
\hline
\end{tabular} 
}
\caption{Inequivalent analytic solutions of the equations \eqref{uni1}-\eqref{uni6}. In the expression of B$2_\pm$, $x\in\left[-\frac{1}{\sqrt{2}}\frac{1}{\sqrt{2}}\right]$ is a free parameter, and $\alpha_\pm(x)\equiv x\,\frac{2x^2-3\pm\sqrt{2(x^2-4)(2x^2-1)}}{2(6x^2+1)}$.}
\label{sol}
\end{table}

%\subsection{Solutions}
The solutions of the fixed point equations (\ref{uni1})-(\ref{uni6}) were determined in \cite{DDL,RPN}, in part analytically and in part numerically. The numerical solutions consist of branches of fixed points extending for $N\leq 2.24421..$ \cite{RPN}, and will not be rediscussed in the present paper, whose main focus is on $N\geq 3$. The analytic solutions are listed in table~\ref{sol}. The table does not include a solution defined only at $N=3$ and called B3 in \cite{RPN}, which was eventually shown in \cite{CPN} to yield the same scattering matrix (\ref{S_tensor}) produced by solution A1 at $N=3$. 

An important feature of equations (\ref{uni1})-(\ref{uni6}) is that for $\rho_4=\rho_5=0$ they reduce to the fixed point equations of the $O(M)$ vector model with $M=M_N$ given by (\ref{M_N}). This implies that the space of solutions of the $RP^{N-1}$ fixed point equations contains the $O(M_N)$ space of solutions as a subspace. The solutions A1, A2 and A3 of table~\ref{sol} indeed coincide with the three solutions of the vector model \cite{DL_ON,sis}. In particular, A2 corresponds to the solution that in the vector model describes the dilute and dense critical branches of the gas of nonintersecting loops \cite{Cardy_book,Nienhuis_review,DeGennes}. Solution A3, on the other hand, possesses $\rho_2$ as a free parameter and describes the line of fixed points which in the $O(2)$ model accounts for the BKT phase. It appears at $N=2$ as it should in view of the known correspondence $RP^1\sim O(2)$, and at $N=-3$ due to $M_{-3}=M_2=2$. Also solutions B1 and B2 of table~\ref{sol} are defined at $N=2$ and contain a free parameter; they correspond to different realizations of the BKT phase in the $RP^{N-1}$ space of parameters.

\subsection{$N\geq 3$}
We are finally left with A1, which is the only solution existing for $N\geq 3$. This means, in particular, that for $N\geq 3$ there are no lines of fixed points at fixed $N$, and then no BKT-like topological transitions yielding quasi-long-range order \cite{DDL}. The possibility of such a transition driven by disclination defects has been debated in numerical studies for long time \cite{CPZ,KZ,FPB,DR,PFBo,FBBP,Tomita,SGR,KS}, and the most recent ones applying the TNR method to the $RP^{2}$ model \cite{UO,BDVVV} confirm our result about its absence. 

Solution A1$_-$ corresponds to $M_N$ free fermions\footnote{The form of crossing and unitarity equation is such that, given a solution, another solution is obtained reversing the sign of all amplitudes. We also recall that the central charge is $1/2$ for a free fermion and 1 for a free boson \cite{DfMS}.}, with central charge $M_N/2$, and is not expected to play a role for the Hamiltonian (\ref{lattice}). We then focus on solution A1$_+$, which completes the $O(M_N)$ subspace of solutions allowed by the $RP^{N-1}$ fixed point equations. In the $O(M_N)$ vector model this solution corresponds to the zero temperature critical point at $M_N>2$ \cite{DL_ON,sis}. As we observed in \cite{DDL,RPN}, this means that the $RP^{N-1}$ model with $N>2$ allows for a zero temperature critical point with the $O(M_N)$ central charge $M_N-1$. Evidence of the realization of this critical point in the square lattice $RP^2$ antiferromagnet has recently been given in \cite{UO} through the determination of the central charge. In three dimensions, where continuous symmetries can break spontaneously, finite temperature criticality in the $O(M_3=5)$ universality class has been identified numerically for the $RP^2$ antiferromagnet on the cubic lattice \cite{Fernandez}, while the transition is first order for the ferromagnet \cite{ZMZ}.  

Solution A1$_+$ corresponds to $M_N$ noninteracting bosons, and always admits the trivial realization with central charge $c=M_N$. Also this trivial realization possesses $O(M_N)$ symmetry, but the large distance behavior is ruled by the $O(M_N)$ realization with minimal central charge $c=M_N-1$. The latter realization is provided by the nonlinear sigma model in which the vector formed by the $M_N$ bosonic fields is constrained to have unit modulus (see e.g. \cite{Cardy_book,Zinn}). Away from criticality, the free and the sigma model realizations have the same particle basis but, of course, different scattering matrices, since in the latter case the constraint introduces interaction. The $O(M)$ sigma model is integrable and its exactly known off-critical scattering matrix \cite{ZZ} can be used to obtain the central charge $M-1$ through the thermodynamic Bethe ansatz or the form factor approach  (see \cite{FZ_O3,BN} for $M=3$). However, when the critical limit of the sigma model scattering matrix is taken sending to infinity the center of mass energy, solution A1$_+$ with $M=M_N$ is recovered. This is how the "asymptotic freedom" of two-dimensional sigma models manifests itself in the scattering framework; in particular, the critical scattering amplitudes do not retain memory of the constraint which reduces the central charge and makes it coincide with the dimensionality of the sigma model target manifold \cite{DL_ON,sis}.  

Extending these considerations we can recognize an additional point. When moving away from criticality, the parameters $\rho_4$ and $\rho_5$, which vanish in the $O(M_N)$-symmetric scattering theory (at and away from criticality), will more generally develop nonzero values. In this case we will have a scattering theory with a different symmetry, but still having A1$_+$ as critical limit\footnote{This point was not sufficiently explored in \cite{DDL,RPN}, where the main effort was absorbed by the derivation of the exact equations and the determination of all the scattering solutions.}. The value $c=2$ measured in \cite{UO} for the central charge of the $T=0$ $RP^2$ ferromagnet is evidence that solution A1$_+$ is also the critical limit of a sigma model with a target manifold of dimension smaller than that of the $O(M_N)$ manifold. This should be the $RP^{N-1}$ manifold with dimension $N-1$, thus setting to $c=N-1$ the central charge of the ferromagnetic fixed point. While this is also the central charge of the $O(N)$ model, the $RP^{N-1}$ and $O(N)$ sigma models have different operator content and represent different universality classes. We already saw that the $RP^{N-1}$ order parameter field is the $N\times N$ traceless symmetric tensor $Q_{ab}(x)$, while the $O(N)$ order parameter field is the vector ${\bf s}(x)$. The local character of the  $\mathbb{Z}_2$ head-tail symmetry of the $RP^{N-1}$ model causes the vanishing of the correlation functions $\langle\cdots{\bf s}(x)\cdots\rangle$, meaning that the $RP^{N-1}$ space of fields does not contain the vector field. This was confirmed by the numerical results of \cite{UO} for $RP^2$, which found $M_3=5$ components for the fundamental field. The energy density field should also differ in the two cases, since $\varepsilon(x)$ appears as the most relevant symmetry-invariant field in the OPE of the order parameter field with itself. One consequence is that the scaling limit of the off-critical $RP^{N-1}$ ferromagnet may be nonintegrable, at variance with $O(N)$.

The numerical results of \cite{UO} for the $RP^2$ central charges are consistent with the conclusion that the $RP^{N-1}$ and $O(M_N)$ sigma models (both having solution A1$_+$ as critical scattering limit) describe the $RP^{N-1}$ ferromagnet and square lattice antiferromagnet, respectively. The reason why ferromagnet and antiferromagnet fall into different universality classes is expected to be the same shown in \cite{AF2} for the three-state Potts model. In the antiferromagnetic case nearest neighbors want to take different values, and the lattice structure matters. The order parameter variable in the square lattice antiferromagnet is defined with an extra sign factor which distinguishes even and odd sublattices ("staggering"). The fields of the continuum then inherit an extra parity related to sublattice exchange. The order parameter has odd parity and cannot produce odd fields in the OPE with itself which determines the energy density field. As a consequence, the field theory of the antiferromagnet cannot contain three-particle vertices that are instead allowed for the ferromagnet\footnote{In their most general form, both Potts and $RP^{N-1}$ Landau-Ginzburg Hamiltonians allow cubic terms which make the transition first order at mean field level \cite{deGP,Wu}.}. The two field theories are different and the absence of three-particle vertices in the antiferromagnet allows the symmetry enhancement displayed by a subspace of the scattering solutions: $\mathbb{S}_3\to U(1)$ for Potts \cite{AF2,DT1}, $RP^{N-1}\to O(M_N)$ for the Hamiltonian\footnote{The symmetry enhancement was proposed in three dimensions in \cite{Fernandez} within a $\phi^4$ description of the $RP^2$ model. In two dimensions all powers of $\phi$ are relevant and normally one needs to deal with exponential fields and their symmetries, as in \cite{AF2,AF1} for the Potts model.} (\ref{lattice}).

\section{$CP^{N-1}$ model}
\label{complex}
The $CP^{N-1}$ lattice model is defined by the Hamiltonian
\begin{equation}
{\cal H}_{CP^{N-1}}=-J\sum_{\langle i,j \rangle}|\mathbf{s}_i \cdot {\mathbf{s}}_j^*|^2,
\label{cpham}
\end{equation}
where $\mathbf{s}_j=(s_{1,j},\ldots,s_{N,j})$ is now a complex $N$-component vector at site $j$ satisfying ${\bf s}_j\cdot{{\bf s}}_j^*=1$. The Hamiltonian  is invariant under global $U(N)$ transformations ($\mathbf{s}_j \to U \mathbf{s}_j$, $U\in U(N)$) and site-dependent $U(1)$ transformations ($\mathbf{s}_j \to e^{i \alpha_j} \mathbf{s}_j $, $\alpha_j\in\mathbb{R}$); these symmetries are represented through the tensorial order parameter variable
\begin{equation}
Q_{ab,i}=s_{a,i}{s}_{b,i}^*-\frac{1}{N}\delta_{ab}\,.
\label{op2}
\end{equation}

The implementation of scale invariant scattering for the $CP^{N-1}$ model closely parallels that seen in the previous section for $RP^{N-1}$ and has been performed in Ref.~\cite{CPN}, to which we refer the reader for the details; here we recall the main points. In the continuum limit the order parameter field $Q_{ab}(x)$ is now a traceless Hermitian tensor. At criticality the massless particle excitations are labeled by a double index $ab$, with $a,b=1,\ldots,N$, and a state containing a particle $ab$ transforms under the $U(N)$ symmetry as 
\begin{equation}
|ab\rangle \goto |a'b'\rangle = \sum_{a,b} U_{a',a} U_{b',b}^* |ab\rangle\,,
\end{equation}
so that the role of the two indices is distinguished by charge conjugation. This is why, while we still have the 11 amplitudes of fig.~\ref{tensor_ampl} parametrized as in (\ref{cr1})-(\ref{cr_last}), the number of possible connections between the terminals is reduced with respect to the $RP^{N-1}$ case, and the scattering matrix takes the form
\begin{equation}
\begin{split}
S_{ab,cd}^{ef,gh} &= S_1\, \delta _{a,d} \delta _{b,c} \delta _{e,h} \delta _{f,g} + S_2 \,\delta _{a,e} \delta _{b,f} \delta _{c,g} \delta _{d,h} + S_3 \,\delta_{a,g} \delta _{b,h} \delta _{c,e} \delta _{d,f} \\
   & \quad + S_4 \left(\delta
   _{a,d} \delta _{b,f} \delta _{c,g} \delta _{e,h}+\delta _{b,c}
   \delta _{a,e} \delta _{d,h} \delta _{f,g}\right) + S_5 \left(\delta
   _{b,c} \delta _{a,g} \delta _{d,f} \delta _{e,h}+\delta _{a,d}
   \delta _{b,h} \delta _{c,e} \delta _{f,g}\right) \\
   & \quad + S_6 \left(\delta
   _{a,e} \delta _{b,h} \delta _{d,f} \delta _{c,g}+\delta _{b,f}
   \delta _{a,g} \delta _{c,e} \delta _{d,h}\right) + S_7 \left(\delta
   _{a,b} \delta _{e,f} \delta _{c,g} \delta _{d,h}+\delta _{c,d}
   \delta _{g,h} \delta _{a,e} \delta _{b,f}\right)\\
   & \quad + S_8 \left(\delta
   _{c,d} \delta _{e,f} \delta _{a,g} \delta _{b,h}+\delta _{a,b}
   \delta _{g,h} \delta _{c,e} \delta _{d,f}\right) + S_9 \big[\delta
   _{e,f} \left(\delta _{a,d} \delta _{b,h} \delta _{c,g}+\delta
   _{b,c} \delta _{a,g} \delta _{d,h}\right)\\
   & \qquad \qquad + \delta _{c,d}
   \left(\delta _{b,f} \delta _{a,g} \delta _{e,h}+\delta _{a,e}
   \delta _{b,h} \delta _{f,g}\right)  \delta _{a,b} \left(\delta
   _{d,f} \delta _{c,g} \delta _{e,h}+\delta _{c,e} \delta _{d,h}
   \delta _{f,g}\right) \\
   & \qquad \qquad + \delta _{g,h} \left(\delta _{a,d} \delta
   _{b,f} \delta _{c,e}+\delta _{b,c} \delta _{a,e} \delta
   _{d,f}\right)\big] + S_{10}\, \delta _{a,b} \delta _{c,d} \delta
   _{e,f} \delta _{g,h} \\
   & \quad+ S_{11} \left(\delta _{a,b} \delta _{c,d}
   \delta _{e,h} \delta _{f,g}+\delta _{e,f} \delta _{g,h} \delta
   _{a,d} \delta _{b,c}\right)\,.
\end{split} \label{Smatrix}
\end{equation}
The decoupling condition of the trace mode ${\cal T}=\sum_a aa$,
\EQ
\mathbb{S}|{\cal T}ab\rangle=S_0|{\cal T}ab\rangle\,, \hspace{1cm}S_0=\pm 1\,,
\label{decoupling2}
\EN
now yields relations which are used to express $S_{i\geq7}$ in terms of $S_{i\leq 6}$ through
\begin{align}
\rho _7 &= \tfrac{1}{N} \left ( S_0 - \rho_2 + \tfrac{2}{N} \big ( 2 \rho_4 \cos \theta  + \rho_5 \big) \right ), \label{decoup1}\\
\rho _8 \cos \psi  &= \tfrac{1}{N} \left ( - \rho_1 \cos \phi + \tfrac{2}{N} \big ( 2 \rho_4 \cos \theta  + \rho _5 \big) \right ), \label{decoup2}\\
\rho _8 \sin \psi &= \tfrac{1}{N}\rho_1 \sin \phi, \label{decoup3}\\
\rho _9 &= -\tfrac{1}{N} \big ( 2 \rho_4 \cos \theta  + \rho_5 \big)  ,\label{decoup4}\\
\rho _{10} &= \tfrac{1}{N^2} \left ( 2 \rho_1 \cos \phi + \rho_2 - S_0  - \tfrac{6}{N} \big ( 2 \rho_4 \cos \theta  + \rho_5 \big) \right ). \label{decoup5}
\end{align}
Unitarity then yields the fixed point equations in the form
\begin{align}
1 &= \rho_1^2 + \rho_2^2 + 2 \rho_4^2 \, , \label{uni1c} \\
0 &= 2 \rho_1 \rho _2 \cos \phi + 2 \rho_4^2 \, , \label{uni2c} \\
0 &= (N^2-1)\rho _1^2 + 2 \rho_1^2 \cos 2 \phi + 2 \rho_1 \rho_2 \cos \phi + 4 \left(N-\tfrac{1}{N}\right) \rho_1 \left ( \rho_4 \cos (\theta -\phi ) + \rho_5 \cos \phi  \right ) \nn
&\quad - \tfrac{4}{N} \rho_1 \rho_4 \cos (\theta +\phi )  + \tfrac{8}{N^2} \rho_4^2 \cos 2 \theta + 2 \left(1 + \tfrac{4}{N^2}\right) \rho_4 \left ( \rho_4 + 2 \rho_5 \cos \theta \right )\nonumber\\
&\quad + 2 \left(1+\tfrac{2}{N^2} \right) \rho_5^2 \, ,  \label{uni3c} \\
0 &= 2 \rho_1 \rho_5 \cos \phi + 2 \rho_2 \rho_4 \cos \theta - \tfrac{4}{N} \rho_4^2 \cos 2 \theta + \left(N - \tfrac{4}{N}\right) \rho_4^2 - \tfrac{8}{N} \rho_4 \rho_5 \cos \theta \nonumber\\
&\quad + \left(N-\tfrac{2}{N}\right) \rho_5^2 \, , \label{uni4c} \\
0 &= 2 \rho_1 \rho_4 \cos (\theta +\phi ) + 2 \rho_2 \rho_5  - \tfrac{4}{N} \rho_4^2 \cos 2 \theta - \tfrac{4}{N} \rho_4^2  + 2 \left(N-\tfrac{4}{N}\right) \rho_4 \rho_5 \cos \theta - \tfrac{2}{N} \rho_5^2 \, , \label{uni5c} \\
0 &= 2 \rho _1 \rho _4 \cos (\theta -\phi )+2 \rho _2 \rho _4 \cos\theta\, . \label{uni6c}
\end{align}

The solutions of these equations were determined in \cite{CPN}, in part analytically and in part numerically. The numerical solutions consist of branches of fixed points extending for $N<2$ and are expected to be relevant for criticality in gases of intersecting loops, examples of which have been discussed in \cite{NSSSO}. The solutions that we determined analytically are listed in table \ref{sol2}. The table does not include solutions defined only for $N=2$ or $N=3$ which were shown in \cite{CPN} to yield the same scattering matrix (\ref{Smatrix}) as solution A1 evaluated at those values of $N$. 

\begin{table}
\centering
%\resizebox{\textwidth}{!}{
\begin{tabular}{|c|c|c|c|c|c|c|c|}
\hline
Solutions & $N$ & $\rho_1$ & $\rho_2$ & $\cos\phi$ & $\rho_4$ & $\rho_5$ & $\cos\theta$\\
\hline\hline
A1$_\pm$ & $\mathbb{R}$ & $0$ & $\pm 1$ & $-$ & $0$ & $0$ & $-$\\
A2$_\pm$ & $[-\sqrt{3},\sqrt{3}]$ & $1$ & $0$ & $\pm\frac{1}{2}\sqrt{3-N^2}$ & $0$ & $0$ & $-$\\
A3$_\pm$ & $\pm \sqrt{3}$ &$\sqrt{1-\rho_2^2}$ & $[-1,1]$ & $0$ & $0$ & $0$ & $-$\\
%B$_\pm$ & $3$ & $\frac{1}{2}$ & $\pm\frac{1}{2}$ & $\mp1$ & $\frac{1}{2}$ & $\rho_2$ & $\pm 1$\\
\hline
\end{tabular}
%}
\caption{Inequivalent analytic solutions of the $CP^{N-1}$ fixed point equations \eqref{uni1c}-\eqref{uni6c}.}
\label{sol2}
\end{table}

It must now be observed that when $\rho_4=\rho_5=0$ equations \eqref{uni1}-\eqref{uni6} reduce to the fixed point equations of the $O(N^2-1)$ model, and that $N^2-1$ is the number of independent real components of the order parameter variable (\ref{op2}). This means that the space of solutions of the fixed point equations of the $CP^{N-1}$ model contains a subspace for which the symmetry is enhanced to $O(N^2-1)$. The solutions A1, A2 and A3 of table~\ref{sol2} all have $\rho_4=\rho_5=0$ and indeed correspond to the three solutions of the $O(M)$ model \cite{DL_ON,sis} with $M=N^2-1$; the fact that $N^2-1=2$ when $N={\pm\sqrt{3}}$ explains the domain of definition of solutions A2 and A3. These conclusions are also consistent with the fact that $CP^1$ corresponds to the Riemann sphere, and then to $O(3)$.

For $N\geq 2$ the fixed point equations possess only solution A1. Taking into account that the fermionic realization A1$_{-}$ should not be not relevant for the Hamiltonian (\ref{cpham}), we are only left with the free bosonic solution A1$_+$, to be associated with zero temperature criticality. Since the theoretical considerations of the previous section for the $RP^{N-1}$ model can be entirely transposed to the present case, we are led to conclude that solution A1$_+$ describes, besides the $O(N^2-1)$ fixed point with central charge $c=N^2-2$, an additional nontrivial fixed point. This will be the fixed point of the $CP^{N-1}$ sigma model, with central charge equal to the dimension of the $CP^{N-1}$ manifold, namely $c=2(N-1)$. While the scattering amplitudes of the $O(N^2-1)$ and $CP^{N-1}$ sigma models have the same critical limit A1$_+$, they will differ away from criticality for $N>2$, since for $CP^{N-1}$ the parameters $\rho_4$ and $\rho_5$ will develop nonvanishing values\footnote{In particular, the $CP^{N-1}$ sigma model is not expected to be exactly solvable away from criticality \cite{GW,BW}. The exception is $N=2$, given that $CP^1\sim O(3)$.}.  

Also the arguments of the previous section about the difference between the $RP^{N-1}$ ferromagnet and antiferromagnet extend to the present case. As a consequence the expectation is that as the temperature goes to zero the Hamiltonian (\ref{cpham}) realizes the $CP^{N-1}$ universality class in the ferromagnetic case, and the $O(N^2-1)$ universality class in the case of the square lattice antiferromagnet. In this respect it must be noted that in three dimensions, where the continuous symmetry can break spontaneously, a finite temperature critical point in the O(8) universality class has been observed in numerical simulations of the antiferromagnetic $CP^2$ model \cite{DPV}.

\section{Conclusion} 
\label{final}
In this paper we disentangled subtleties hidden in the space of solutions of the exact fixed point equations of the two-dimensional $RP^{N-1}$ and $CP^{N-1}$ models that we had determined in \cite{DDL,RPN,CPN} using scale invariant scattering theory. In particular, we explained why the unique relevant solution A1$_+$ of  critical $RP^{N-1}$ scattering for $N\geq 3$ actually corresponds to two different universality classes with the same number $M_N=\frac{1}{2}N(N+1)-1$ of order parameter components. These are the $O(M_N)$ universality class with central charge $M_N-1$, and the $RP^{N-1}$ universality class with central charge $N-1$. While the scattering amplitudes for the two universality classes will differ away from criticality, they have the same critical limit A1$_+$ due to the asymptotic freedom of two-dimensional nonlinear sigma models. When the temperature goes to zero, the $RP^{N-1}$ lattice Hamiltonian (\ref{lattice}) realizes the $RP^{N-1}$ universality class for ferromagnetic interaction, and the $O(M_N)$ universality class for antiferromagnetic interaction on bipartite lattices. The symmetry enhancement in the antiferromagnetic case is due to a suppression of three-particle vertices analogous to that shown in \cite{AF2} for the three-state Potts model.

We also discussed how a similar pattern is expected in the $CP^{N-1}$ model (\ref{cpham}), with the unique scattering solution for $N>2$ corresponding to the critical limit of both the $O(N^2-1)$ universality class with central charge $N^2-2$ and the $CP^{N-1}$ universality class with central charge $2(N-1)$. 

The fact that a single scale invariant scattering solution may correspond to different renormalization group fixed points was already known for the $q$-state Potts model \cite{DT1}. In the $RP^{N-1}$ and $CP^{N-1}$ models the mechanism is made even more subtle by the fact that it simultaneously accounts for symmetry enhancement in the antiferromagnetic case for generic values of $N$.

%\end{document}
%\newpage
%\vspace{1cm} \noindent \textbf{Acknowledgments.} 

\end{document}